\documentstyle[aps,prl,multicol,epsf]{revtex}

\begin{document}
\draft
\title{ Holonomic Quantum Computation}
\author{Paolo Zanardi $^{1,2}$ and Mario Rasetti $^{1,3}$ }
\address{ $^1$ Istituto Nazionale Fisica della Materia (INFM)\\
$^2$ Institute for Scientific Interchange Foundation,
Villa Gualino, Viale Settimio Severo 65, I-10133 Torino, Italy\\
$^3$ Dipartimento di Fisica, Politecnico di Torino,
Corso Duca degli Abruzzi 24, I-10129 Torino, Italy}
\maketitle
\begin{abstract}
{ We show that  the notion of generalized Berry phase i.e., non-abelian holonomy,
can be used for enabling
 quantum computation.
The computational space is realized by a  $n$-fold degenerate eigenspace
of a family of Hamiltonians parametrized by a manifold $\cal M$.
The point of $\cal M$ represents classical configuration
of control fields and, for multi-partite systems, couplings between subsystem.
Adiabatic loops in the control $\cal M$ induce non trivial unitary transformations on the computational
space. For a generic system it is shown that this mechanism allows for universal quantum computation
by composing a generic pair of loops  in $\cal M.$
}
\end{abstract}
\date{today}
\pacs{PACS numbers: 03.67.Lx, 03.65.Bz}
\maketitle
\begin{multicols}{2}
\narrowtext
In this paper we shall  speculate about  a novel potential application
of non-abelian geometric phases ({\em holonomies}) to Quantum Computation \cite{QC}.
Ever since their discovery geometric phases in quantum theory have been considered
a deep and   fascinating subject \cite{SHWI}.
This is due on the one hand to their unexpected and ubiquitous 
role in many physical systems, on the other hand
to the elegant formulation they admit in terms
of concepts borrowed from differential-geometry and topology \cite{NAK}.
Furthermore the existence  of analog   geometric  terms
associated with {\em non-abelian} groups e.g., $U(N)$ \cite{WIZE}  showed 
how many of the notions  developed in (non-abelian) gauge theory  
have  a scope that extends far beyond the study of fundamental interactions.

We shall show how by  encoding quantum information in one of the  eigenspaces
of a  degenerate Hamiltonian $H$ one can in principle achieve the full quantum computational power  
by using holonomies only.
These holonomic computations are realized by  moving along loops in a suitable space $\cal M$
of control parameters labelling
the family of Hamiltonians to which $H$ belongs.
Attached to each point $\lambda\in{\cal M}$ there is a quantum code,
and this  bundle of codes is endowed with a non-trivial  global topology described by a non-abelian gauge field 
potential $A$.
For generic $A$ the associated holonomies will allow for universal quantum computing. 
In a sense the ideas presented here suggest that gauge fields might
play a  role also in the arena of information processing.

In the very same  way of classical information processing,  general quantum computations
are realized by networks of   elementary bulding blocks.
More specifically the dynamics is obtained by { switching} on and off {\em gate } Hamiltonians $\{H_l\}_{l=1}^g.$
If in this way any unitary over the state-space  
can be approximated arbitrarily well, the set
of gates $U_l:=e^{i\,H_l}$ is termed {\em universal} \cite{UG}.
A (universal) quantum computer  is defined by the state-space ${\cal H}\cong {\bf{C}}^N$
for data encoding
and by a (universal) set of quantum gates. A quantum algorithm  consists of
the given 
computation $U(T)$ that acts on the   quantum state $|\psi\rangle_{in}$ encoding
 initial data, its realization as a network
of basic gates, along with a (measurement) prescription for extracting
the relevant information from $|\psi\rangle_{out}.$
For the aims of this paper it is worthwile to reformulate this
setup in a geometrical fashion.

Any quantum evolution $U(T)={\bf{T}}\,\exp[-i\,\int_0^T dt H(t)]$ can be  associated with a path 
in a   space whose points  describe  the  configurations
of  suitable ``control fields'' $\lambda,$ on which the Hamiltonian depends.
Indeed
any Hamiltonian in ${\bf{C}}^N$ can be written as 
$H_\lambda=i\,\sum_{l=1}^{N^2} \Phi_l(\lambda) \,\Gamma_l\,(\Phi_l\in{\bf{R}})$
where  the $\Gamma_l$'s are   a basis of the 
space of anti-hermitean matrices i.e.,
they are the generators of the Lie algebra  $u(N).$
The control parameter space $\cal M$ is
 a  manifold  over which is defined a smooth map $\Phi$
to  $u(N).$ 
If one is able
to drive the control field configuration $\lambda\in{\cal M}$ through a
(smooth) path $\gamma\colon [0,\,T]\rightarrow {\cal M}$
then a  family $H(t) :=H_{\gamma(t)}$ is defined along with the associated unitary $U_\gamma.$
Conversely any smooth family $H(t)$ defines a path in ${\cal M}={\bf{R}}^{N^2}.$ 
Resorting to this language Quantum Computation can be described as the
experimenter's capability of generating  
a small set of $\{\gamma_i\}_{i=1}^g$ of basic paths such that
sequences of the corresponding $U_{\gamma_i}$'s approximate
with arbitrary good accuracy any unitary transformation on the quantum state-space.  
It is important to stress the, obvious, fact
that the path generation is achieved through a  classical control process.

{\em Non-Abelian  Holonomies}-- %%%%%%%%%%%%%%%%%%%%%%%%%%%%%%%%%%%%%%%%%%%%%%%%%%%%
In the  situation which we are interested in,  
one deals with  
 $\gamma$'s that are  {\em loops} in $\cal M$ i.e., $\gamma(T)=\gamma(0),$
and with a   family of  Hamiltonians
$\{ H_\lambda \}_{\lambda\in {\cal M}}$
with same degeneracy 
structure i.e., no-level crossing.
In the general case  $H_\lambda=H_{\gamma(t)}$ has $R$ different eigenvalues $\{ \varepsilon_i\}_{i=1}^R$
with degeneracies $\{n_i\}.$
If $\Pi_i(\lambda)$ denotes the projector over the  eigenspace
${\cal H}_i(\lambda)  :=\mbox{span}\,\{|\psi^\alpha_{i}(\lambda)\rangle\}_{\alpha=1}^{n_i},$
 of $H_\lambda,$ one has the spectral $\lambda$-dependent resolution 
$ H_\lambda=\sum_{i=1}^R \varepsilon_i(\lambda) \, \Pi_i(\lambda).$

The state vector evolves according the time-dependent Schroedinger equation 
$i\,\partial_t|\psi(t)\rangle= H_{\gamma(t)}\,|\psi(t)\rangle.$
We shall restrict ourselves 
to the case in which the loop $\gamma$'s are  {\em adiabatic}
i.e., $\hbar{\dot\gamma}/\gamma\ll \mbox{min}_{i\neq j} |\varepsilon_j-\varepsilon_i|.$
Then it is well known that any initial preparation  $|\psi_0\rangle\in{\cal H}$
will be mapped, after the period $T,$ onto:
%\begin{equation}
$
|\psi(T)\rangle=
U(T)\,|\psi_0\rangle,\, U(T)=\oplus_{l=1}^R e^{i\,\phi_l(T)}\,\Gamma_{A_l}(\gamma),
$
%\end{equation}
where, $\phi_l(T):=\int_0^T d\tau\, \varepsilon_l(\lambda_\tau),$ is the dynamical phase and  
\begin{equation}
 \Gamma_{A_l}(\gamma) :={\bf{P}}\exp \int_\gamma A_{l} \in U(n_l),
\quad(l=1,\ldots,R)
\label{hol}
\end{equation}
is called the {\em holonomy} associated with the loop $\gamma,$
(here ${\bf{P}}$ denotes path ordering).
In particular  when  $|\psi_0\rangle\in {\cal H}_l$ the final state 
 belongs to the {\em same} eigenspace.
In the following we will drop dynamical phases and focus on the geometrical contribution
(\ref{hol}).
For $n=1$ this  term  is nothing but the celebrated  Berry phase, and $A$
is the so-called Bott-Chern connection \cite{BOCH}. For $n_l>1$
the holonomy $\Gamma_{A_l}(\gamma)$ 
is sometimes referred to as {\em non-abelian} geometric phase \cite{WIZE}.
The matrix-valued form $A_l$ appearing in Eq. (\ref{hol}) is known as the
{\em adiabatic connection}
and it is given by $A_{l} = \Pi_l(\lambda)\,d\,\Pi_l(\lambda)=\sum_\mu
A_{l,\mu}\,d\lambda_\mu,$ where \cite{SHWI}
\begin{equation}
(A_{l,\mu})^{\alpha\beta}:= \langle\psi_{l}^\alpha(\lambda)|
\,{\partial}/{\partial\lambda^\mu}\,
|\psi_{}^\beta(\lambda)\rangle
\label{conn}
\end{equation}
$(\lambda_\mu)_{\mu=1}^d$ local coordinates on ${\cal M}.$
The $A_l$'s are a non-abelian gauge  potentials that allow for parallel transport
of vectors over $\cal M.$
Indeed the linear mapping (\ref{hol}) of the fiber ${\cal H}_l$ onto itself 
is nothing but the parallel transport of the vector $|\psi_0\rangle$ associated with the 
connection form $A_l.$

In view of the crucial role played
by degeneracy, before moving to the main part of the paper, 
 we briefly discuss
this issue in a geometric fashion  by considering the space of Hamiltonians $H$ of a
quantum state-space ${\cal H}\cong{\bf{C}}^N$.

The control  manifold is mapped by $\Phi$ onto   a set of Hamiltonians iso-degenerate with
$H=H_{\gamma(0)}$.
Locally one has 
$\Phi({\cal M})\cong {\cal O}(H)\times ({\bf{R}}^R-\Delta_R),$
 where
$\Delta_R:=\{x\in{\bf{R}}^R\,\colon\, i\neq j\Rightarrow x_i\neq x_j\},$ and 
 ${\cal O}(H):=\{ X\,H\,X^\dagger\,/\,X\in U(N)\}$ is the orbit 
of $H$ under  the (adjont)
action of $U(N).$
Indeed any pair of isospectral Hamiltonians belongs to  ${\cal O}(H),$ moreover 
once the orbit is given  one has still $R$ degrees of freedom
(the different eigenvalues) for getting the whole manifold of Hamiltonian with fixed degeneracy structure.
By factoring out
the  the symmetry group of $H,$
one finds
 \begin{equation}
{\cal O}(H):=\frac{U(N)}{ U(n_1)\times\cdots\times U(n_R)},
\label{flag}
\end{equation}
From eq. (\ref{flag}) it stems that dimension of this manifold    reach its maximum (minimun)
for the non (maximally) degenerate case $R=N$ ($R=1$):
$d_{max}= N\,(N-1)+N=N^2$ ($d_{min}=0+1=1$).
This means that the set  of non-degenerate Hamiltonians is an {\em open}
submanifold of ${\bf{R}}^{N^2},$  expressing the well-known fact that degeneracy
-- due to the symmetry constraints
that it involves -- is a singular case, while non-degeneracy is the generic  one.
Indeed if one slightly perturbs a non-degenerate Hamiltonian $H$
the resulting operator  is, generically, still non-degenerate.

{\em Universal Computation} --%%%%%%%%%%%%%%%%%%%%%%%%%%%%%%%%%%%%%%%%%%
The above considerations make clear that the degeneracy requirement for the existence
of non-abelian holonomies is rather stringent from  a purely geometrical point of view.
On the other hand quite often the physics of the  systems under concern provides
the required symmetries for having (large) degenerate eigenspaces.
Notice that discrete symmetries, like charge conjugation
and rotational invariance are rather generic in many-body systems.
For example non-abelian holonomies have been recently shown
to play a role in the $SO(5)$ theory of superconductivity \cite{ZHA}.
 
In the following we will take degeneracy for granted and we will
fix our attention to a given $n$-dimensional eigenspace $\cal C$ of $H.$
The state-vectors in $\cal C$ will be our quantum codewords,
and $\cal C$ will be referred to as the {\em code}.
Clearly the optimal choice is to take the code to be
the largest eigenspace of $H.$
Our aim is to perform  as many as possible unitary transformations i.e., {\em computations, }
 over the code resorting  only on   the non-abelian holonomies (\ref{hol}) generated by adiabatic loops
in $\cal M.$
A first crucial  question is:

{\em
How many transformations can be obtained, by eq. (\ref{hol}), as $\gamma$ varies
over the space of loops in $\cal M$?
 } 

To address this point let us begin by considering
the  properties of the holonomy map $\Gamma_A.$ 
On the loop space (we set $T=1$) 
\begin{equation}
L_{\lambda_0}:=\{\gamma\colon [0,\,1]\mapsto {\cal M}\,/\,
\gamma(0)=\gamma(1)=\lambda_0\}
\end{equation}
over a  point $\lambda_0\in{\cal M},$ 
there exists
a composition  law for loop [i.e., $(\gamma_2\cdot \gamma_1)(t)=\theta(
\frac{1}{2} -t)\,\gamma_1(2\,t)+
\theta(t-\frac{1}{2})\,\gamma_1(2t-1)$] and a unity element $\gamma_0(t)=\lambda_0,\,t\in[0,\,1].$
The basic property of map $\Gamma_A\colon L_{\lambda_0}\mapsto
U(n)$
are easily derived from eq. (\ref{hol}):
i) $\Gamma_A(\gamma_2\cdot\gamma_1)=\Gamma_A(\gamma_2)\,\Gamma_A(\gamma_1)$;
ii) $\Gamma_A(\gamma_0)=\openone$;
moreover, by denoting with $\gamma^{-1}$ the loop $t\mapsto \gamma(1-t),$ one has
iii) $\Gamma_A(\gamma^{-1})=\Gamma_A^{-1}.$
This means that by composing loops in $\cal M$ one obtains 
a unitary evolution that is the product of the evolutions associated with the individual loops
and that  staying at rest in the parameter space correspond to no evolution at all.
Finally iii) tells us that for getting the  time-reversed evolution one has
simply to travel along $\gamma$ with the  opposite orientation.
Another noteworthy property of $\Gamma_A$ -- on which its geometric nature
is based -- is its invariance under reparametrizations:
$\Gamma_A(\gamma\circ \varphi)=\Gamma_A(\gamma),$ where $ \varphi$ is any
diffeomorphism of $[0,\,1].$
Physically this means that the evolution map
 -- as long as adiabaticity holds --
does not depend on the rate at which $\gamma$ is travelled but just on its
geometry.
This property is quite non-trivial and, obviously,  does not hold
for general time-dependent quantum evolutions.

From i)--iii) it follows immediately that
the set $\mbox{Hol}(A):=\Gamma_A(L_{\lambda_0})$ is a {\em subgroup} of $U(n)$ known
as the {\em holonomy group} of the connection $A.$ 
Notice that the distinguished  point $\lambda_0$
is not crucial, in that $\Gamma_A(L_{\lambda_0})\cong\Gamma_A(L_{\lambda^\prime_0})$
provided $\lambda_0$ and $\lambda_0^\prime$ can be conncetd by a smooth path.
When ${\mbox{Hol}(A)}=U(n)$, the connection $A$ is called {\em irreducible}.
To our aims  the key observation is that irreducibility is the {\em generic} situation.
This result can be stated geometrically by saying that in the space of connections over $\cal M
,$ the
irreducible ones are an open dense set.
The condition of irreducibility can be stated in terms of the  {\em curvature} $2$-form of the connection
$F=\sum_{\mu\nu}F_{\mu\nu}\,dx^\mu\wedge dx^\nu$ where
\begin{equation}
F_{\mu\nu} =\partial_\nu A_\mu-\partial_\mu A_\nu - [A_\mu,\,A_\nu].
\label{F}
\end{equation}
If the $F_{\mu\nu}$'s linearly span the whole Lie algebra $u(n)$, then $A$ is irreducible \cite{NAK}.
It follows that  in the generic case adiabatic connections will provide a mean for realizing
universal  quantum computation  over ${\cal C}.$ 
For any chosen unitary transformation $U$ over the code
there exists a path $\gamma$ in $\cal M$ such that $\|\Gamma_A(\gamma)-U\|
\le \epsilon ,$ with $\epsilon$  arbitrarily small.
Therefore  any computation on the   code $\cal C$
can be realized by driving the control  fields configuration $\lambda$ along
closed paths $\gamma$ in the control manifold $\cal M.$

Now we show that the connections associated with non abelian geometric phases
are {\em actually} irreducible.
For simplicity in eq. (\ref{flag})
we set $R=2,\,n_1=1,\,n_2=N-1$
obtaining the $N-1$-dimensional complex projective space
\begin{eqnarray}
{\cal O}(H_0)
\cong \frac{ U(N)}{U(N-1)\times U(1)} \cong\frac{SU(N)}{U(N-1)}
\cong
{\bf{CP}}^{N-1}.
\end{eqnarray}
The orbit ${\cal O}(H_0)$ of $H_0 \equiv H_{\lambda_0}$ coincides
with the manifold of pure states over ${\bf{C}}^{N}.$
When $N=2$  one recovers the  original Berry-Simon case,  
$H_{BS}={\bf{B}}\cdot {\bf{S}}, ({\bf{S}}:=(\sigma_x,\,\sigma_y,\,\sigma_z),\,
{\bf{B}}\in S^2\cong{\bf{C P}}^1),$
for a  spin $\frac{1}{2}$ particle in an external magnetic field ${\bf{B}}$.
Here Hol$\,(A_{BS})=\{e^{i\,S_{\gamma}}\}_\gamma\cong U(1),$ where
$S_\gamma$ is the area enclosed by the loop $\gamma$ in the sphere $S^2.$
Of course this  case, being abelian, has no computational meaning, 
nevertheless it shows how controllable loops  in an external field manifold (the ${\bf{B}}$-space)
can be used for generating quantum phases.

For the characterization of the holonomy group
we observe first of all that one
 can identify the control manifold with orbit ${\cal O}.$
Technically this is due to the fact that the bundle of $N-1$-dimensional ``codes''
over ${\cal M}$ is   vector bundle with structure group  $U(N-1).$
The associated $U(N-1)$-principal bundle is the pull back, through $\Phi,$
of $\pi\colon U(N)/U(1)\mapsto {\bf{CP}}^{N-1}.$
The result follows being the latter an universal classifying bundle \cite{Ali}.

For general $N$ the points of ${\bf{C P}}^{N-1}$ are parametrized by the transformations
${\cal{U}}({\bf{z}}):= {\bf{P}}\,\prod_{\alpha=1}^{N-1} U_\alpha(z_\alpha),$ where
%\begin{equation}
$
U_\alpha(z_\alpha)
:=\exp ( z_\alpha\,|\alpha\rangle\langle N|-\mbox{h.c.}).
$
%\end{equation}
The relevant projectors are given by $\Pi_{{\bf{z}}}={\cal{U}}({\bf{z}})\Pi\,
{\cal{U}}({\bf{z}})^\dagger,$ where $\Pi$ is the projector over the first $N-1$
degenerate eigenstates.
By using def. (\ref{F}) and setting $z_\alpha=z^0_\alpha+i\,z^1_\alpha,$
one checks that at ${\bf{z}}=0$ the  components of the curvature
are given by 
\begin{equation}
F_{z_\alpha^n, z_\beta^m}(0) =\Pi\,[\frac{\partial U_\alpha}
{\partial z^n_{\alpha}},\,\frac{\partial U_\beta}
{\partial z_{\beta}^m}]\,\Pi|_{{\bf{z}}=0},
\end{equation}
with $\alpha,\beta=1,\ldots,N-1, \, ;\, m,n=0,1.$

Since
${\partial U_\alpha}/
{\partial z^n_{\alpha}}=i^n\,(|\alpha\rangle\langle N|-(-1)^n\,|N\rangle\langle \alpha|),
$
one finds
\begin{equation}
F_{z_\alpha^n, z_\beta^m}(0)=i^{m+n}\,[
(-1)^{n}|\beta\rangle\langle\alpha|-
(-1)^m\,|\alpha\rangle\langle\beta|].
\end{equation}
From this expression it follows that
components of $F$ span  the whole $u(N-1).$
As remarked earlier, this result does not depend on the specific point chosen, therefore
this example is irreducible i.e.,  Hol$(A)\cong U(N-1).$
The general case (\ref{flag}) can be worked out along similar  lines
it turns out to be irreducible as well.
Notice how, for generating control  loops for $N$ qubits, 
one needs to control $2^{N+1}$ real parameters
instead of the $2^{2\,N}$ ones necessary for labelling a generic Hamiltonian.

For practical purposes is relevant
the  question:

{\em
How many loops should an experimenter be able to generate
for getting the whole holonomy group?
}

An existential answer is given below by using arguments close to the ones
of ref. \cite{LLO}. As 
the non-trivial topology associated with the irreducible gauge-field $A$
allows to map the loop ``alphabet'' densely into the group of unitaries over
the code, we have the 

{\em Proposition}
Two generic loops $\gamma_i\,(i=1,2)$  generate a universal set of gates over $\cal C.$

{\em Proof.} It is known   that  two generic unitaries $U_1$ and $U_2$ belonging to a subgroup $\cal G$
of  $U(N)$ generate, by composition, a subgroup $G$ dense in  $\cal G$ \cite{LLO}.
In particular if  ${\cal G}=U(N)$ the $U_i$'s are a universal set of gates. 
Formally: let $U_{\pm\alpha}:= U^{\pm 1}_\alpha \,(\alpha=0,\pm 1, \pm 2) \,
;\, U_0:={\bf{\openone}},$
then the set $G$ of transformations obtainable by composing the $U_i$'s (along with their inverses)
is given by  $U_f:={\bf{P}}\, \prod_{p \in {\bf{N}} } U_{f(p)},$ 
where $f$ is a map from the natural numbers ${\bf{N}}$
to the set $\{0,\pm 1, \pm 2\}$ nonvanishing only for finitely many $p$'s.

From the basic 
 relation i) it follows that the transformations $U_f$ are  generated by composing loops in $L_{\lambda_0}:$
\begin{equation}
U_f=\Gamma_A(\gamma_f),\quad \gamma_f:={\bf{P}} \prod_{p \in {\bf{N}} }\gamma_{f(p)}.
\end{equation}
We set $\{U_i:=\Gamma_A(\gamma_i)\}_{i=1}^2,$ then $\overline{G}= \mbox{Hol}(A)=U(N),$
the latter relation follows from irreducibility of $A.$
$\hfill\Box$

Of course this result  does not provide an explicit
recipe for obtaining the desired transformations,
nevertheless it is conceptually quite remarkable.
It shows that, even though adiabatic holonomies
are a very special class of quantum evolutions, they
still provide  the full computational power
for processing quantum information.

On the other hand our result is not completely surprising.
Indeed it is important to stress that the  parameters $\lambda$,
for a  multi-partite system, will  contain in general external fields
as well as couplings between sub-systems.
For example if ${\cal H}={\bf{C}}^2\otimes{\bf{C}}^2$ is two-qubits space
a possible basis for $u(4)$ is given by $i\,\sigma_\mu\otimes\sigma_\nu,$
where $\sigma_0:=\openone$ and $\{\sigma_i\}_{i=1}^3$ are the Pauli matrices.
Then for $ij\neq 0$ the $\Gamma_{ij}$'s describe non-trivial interactions between the two qubits,
while for $ij=0$ the corresponding generators are single qubit operators.
Only the control fields associated with these latter $\Gamma_{ij}$'s can be properly interpreted
as external fields while the others generate true entanglement among subsystems.
Moreover, quite often, the parameters $\lambda$ are indeed 
 quantum degrees of freedom,
which are considered frozen in view of the adiabatic
decoupling \cite{SHWI}.
In this case the generation of loops $\gamma$
is on {\em its own} a problem of quantum control. 

So far we have been concerned just with existential issues
of unitary evolutions. In the following we shall briefly address
the associated problem of computational complexity.
A detailed discussion of this  point is given elsewhere \cite{ISI}.
It is widely recognized that a crucial ingredient that provides
quantum computing with its additional power is {\em entaglement}.
This means that the computational state-space has to be multi-partite
e.g., ${\cal H}=({\bf{C}}^2)^{\otimes\,N},$
and the computations are, efficiently, obtained by composing local gates
that act non trivially over a couple of subsystems at most \cite{UG}.
In general the degenerate eigenspaces in which we perform
our holonomic computations do not have any preferred tensor product structure.
Once one of these structures has been chosen over a $N$-qubit code ${\cal C}$ 
i.e., an  isomorphism $\varphi\colon{\cal C}\mapsto ({\bf{C}}^2)^{\otimes\,N}$ has been
selected, any unitary transformation over ${\cal U}$ can  written as a suitable sequence
of CNOT's and single qubit transformations.
In the  ${\bf{CP}}^N$  model discussed above 
it can be proven, by explicit computations, that one can contructively get 
any single-qubit and two-qubit gate
as well by composing elementary holonomic loops   
restricted with suitable $2$-dimensional manifolds.
The point, bearing on the complexity issue, is that the  number of such elementary 
loops scales exponentially as a function of the qubit number. 

A possible way out is given by considering 
a system that is multipartite from the outset and 
a special form of the Hamiltonian
family $H(\lambda).$
The latter is given by a sum, over all the possible pairs $(i,\,j)$ of subsystems,
of Hamiltonian families $\{H(\mu_{ij})\}.$ 
Suppose that  the dependence on the local control 
parameters $\mu_{ij}$ is such that one can holonomically generate any transformation
on a two-qubit subspace ${\cal C}_{ij}\subset{\cal H}_i\otimes{\cal H}_j$
e.g., a $U(8)/U(4)\times U(4)$-model, then one can {\em efficiently}
generate any unitary over the computational subspace $\otimes_{(i,\,j)} {\cal C}_{ij}$
by using holonomies only \cite{ISI}.

%%%%%%%%%%%%%%%%%%%%%%%%%%%%%%%%%%%%%%%%%%%%%%%%%%%%%%%%%%%

{\em An example}-- %%%%%%%%%%%%%%%%%%%%%%%%%%%%%%%%% 
Let ${\cal H}:=\mbox{span}\,\{|n\rangle\}_{n\in{\bf{N}}}$ be 
the Fock space of a single bosonic mode, $H_0=\hbar\omega\, n (n-1) \,(n:=a^\dagger a,\,[a,\,a^\dagger]=1).$ 
Hamiltonians of this kind can arise in quantum optics
when one condider higher order non-linearities.
By construction the space ${\cal C}=\mbox{span}\,\{ |0\rangle,\,|1\rangle\}$
is a two-fold degenerate  eigenspace of $H_0$ i.e., $H_0\,{\cal C}=0.$
Consider the two-parameter isospectral family of Hamiltonians 
$H_{\lambda\mu}:=U_{\lambda\mu}\,H_0\, \,U_{\lambda\mu}^\dagger,
\,(\lambda,\mu\in{\bf{C}})$
where
\begin{equation}
U_{\lambda, \mu}:=
\exp(\lambda\,a^\dagger-\bar\lambda\,a)\,\exp(\mu\,a^{2\dagger}-\bar\mu\,a^2).
\end{equation}
The  first (second)  factor in this equation is nothing but 
the unitary transformation from the Fock vacuum $|0\rangle$ to the
familiar coherent (squezeed) state basis.
If $\Pi$ denotes the projector over the degenerate eigenspace of $H_0,$ 
one gets $A= \Pi\,U_{\lambda\mu}^{-1}\,d U_{\lambda\mu}\,\Pi=
 A_\lambda d\lambda +A_\mu d\mu -\mbox{h.c.},$
where (at $\lambda=\mu=0$)
$
A_\lambda :=-\Pi\,a^\dagger\,\Pi, \quad A_\mu :=-\Pi\,
a^{2\dagger}\,\Pi.$
From this relations the explicit matrix form
of $A$ can be immediately computed and irreducibility for the single-qubit 
space $\cal C$
verified.

This example, at the formal level,  can be easily generalized.
i) Choose an Hamiltonian $H$ belonging to  a representation $\rho$ of some  dynamical (Lie) algebra $\cal A,$
ii) Build a  $k$-fold degenerate $H_f:=f(H),$ iii) Consider the orbit of ${\cal O}(H_f)=f({\cal O}(H))$
 under
the inner automorphisms of $\cal A.$
In point ii)  $f$ is a smooth real-valued map
such  that $f(\varepsilon_i)=E,\,(i=1,\ldots,k),$ with the  $\varepsilon_i$'s belong to some subset
of the spectrum of $H.$
In the present case one has  ${\cal A}:=\{a,\,a^\dagger,\,a^2,\,a^{\dagger\,2},\, n:=a^\dagger\,a,\,\openone\},$
$f(z)=z\,(z-1),$ and $\rho$ is the  bosonic Fock representation.

{\em Conclusions}-- %%%%%%%%%%%%%%%%%%%%%%%%%%%%%%%%%%%%
In this paper we have shown how the notion of non-abelian holonomy (generalized Berry phase)
might in principle provide a a novel  way for implementing universal quantum computation. 
The quantum space (the code) for encoding information is realized by a degenerate eigenspace
of an Hamiltonian belonging of a smooth iso-degenerate   family parametrized 
by points of  a control manifold $\cal M$. 
The computational bundle of eigenspaces over $\cal M$ is endowed by a non-trivial holonomy 
associated with  a generalized Berry connection $A.$
Loops in ${\cal M}$ induce unitary transformations over the  code attacched to a distinguished point 
$\lambda_0\in{\cal M}.$
We have shown that, in the generic i.e., irreducible, case
universal quantum computation can then be realized by composing in all possible ways 
a pair of adiabatic loops.

The required capability of generating
loops by changing   coupling constants, along with the necessity of  large degenerate eigenspace,
makes evident that from the experimental point of view the scheme we are analysing 
is exceptionally demanding like any other proposal for quantum computing. 
However we think that the connection between  a differential-geometric    
concept like that of non-abelian holonomy and the general problematic
of quantum information processing is non-trivial and quite intriguing. 
The individuation of promising physical systems for implementing
the ``gauge-theoretic'' quantum computer we have been discussing in this paper
is still an open problem that will require a  deal of 
further investigations.

The authors thank H. Barnum, G. Segre and L. Faoro for discussions,
A. Uhlmann for useful correspondence.
P.Z. is supported by Elsag, a Finmeccanica Company.
%%%%%%%%%%%%%%%%%%%%%%%%%%
%%%%%%%%%%%%%%%%%%%%%%%%%%%%%%%

\end{multicols}

\begin{references}
\bibitem{QC} For reviews, see D.P. DiVincenzo, {\sl Science} {\bf
270}, 255 (1995); A. Steane, Rep. Prog. Phys. {\bf 61}, 117 (1998)
\bibitem{SHWI} For a review see, {\em Geometric Phases in Physics}, A. Shapere and F. Wilczek, Eds.
World Scientific, 1989
\bibitem{NAK} M. Nakahara, {\em Geometry, Topology and Physics}, IOP Publishing Ltd., 1990
\bibitem{WIZE} F. Wilczek and A. Zee., Phys. Rev. Lett. {\bf {52}}, 2111 (1984)
\bibitem{UG} D. Deutsch, A. Barenco and A. Ekert, Proc. R. Soc. London { A}, {\bf 449}, 669 (19
95); D.P. Di Vincenzo, Phys. Rev. A, {\bf 50}, 1015 (1995)
\bibitem{BOCH} R. Bott, and S.S. Chern, {\sl Acta Math.} {\bf 114}, 71 (1985) 
\bibitem{ZHA} E. Demler and S.C. Zhang,
LANL e-print archive {\tt cond-mat/9805404}
\bibitem{Ali} A. Mostafazadeh, J. Math. Phys. {\bf 37},  1218 (1996)
\bibitem{LLO} S. Lloyd, Phys. Rev. Lett. {\bf {75}}, 346 (1995)
\bibitem{ISI} J. Pachos et al. Phys. Rev. A (To be published) 

\end{references}
\end{document}